\documentclass[aps,prl,preprint,superscriptaddress]{revtex4-1}
\usepackage[pdftex]{graphicx}

\begin{document}


\title{$\alpha$-cluster ANCs for nuclear astrophysics}


\author{M.L. Avila}
\email[]{mavila@anl.gov}
\altaffiliation{Physics Division, Argonne National Laboratory, Argonne IL 60439, USA}
\affiliation{Department of Physics, Florida State University, Tallahassee, FL 32306, USA}

\author{G.V. Rogachev}
\email{rogachev@tamu.edu}
\affiliation{Department of Physics\&Astronomy and Cyclotron Institute, Texas A\&M University, College Station, TX 77843, USA}

\author{E. Koshchiy}
\affiliation{Department of Physics\&Astronomy and Cyclotron Institute, Texas A\&M University, College Station, TX 77843, USA}

\author{L.T. Baby}
\affiliation{Department of Physics, Florida State University, Tallahassee, FL 32306, USA}

\author{J. Belarge}
\affiliation{Department of Physics, Florida State University, Tallahassee, FL 32306, USA}

\author{K.W. Kemper}
\affiliation{Department of Physics, Florida State University, Tallahassee, FL 32306, USA}

\author{A.N. Kuchera}
\altaffiliation{National Superconducting Cyclotron Laboratory, Michigan State University, East Lansing, MI 48824, USA}
\affiliation{Department of Physics, Florida State University, Tallahassee, FL 32306, USA}

\author{D. Santiago-Gonzalez}
\altaffiliation{Department of Physics and Astronomy, Louisiana State University, Baton Rouge, LA 70803, USA}
\affiliation{Department of Physics, Florida State University, Tallahassee, FL 32306, USA}


\date{\today}

\begin{abstract}
\begin{description}
\item[Background] 
Many important $\alpha$-particle induced reactions for nuclear astrophysics may only be measured
using indirect techniques due to small cross sections at the energy of interest.
One of such indirect technique, is to determine the Asymptotic Normalization Coefficients 
(ANC) for near threshold resonances extracted from sub-Coulomb $\alpha$-transfer reactions. 
This approach provides a very valuable tool for studies of astrophysically important reaction
rates since the results are practically model independent. However, the validity of the method 
has not been directly verified.

\item[Purpose]
The aim of this letter is to verify the technique using the 
$^{16}$O($^6$Li,$d$)$^{20}$Ne reaction as a benchmark. 
The $^{20}$Ne nucleus has a well known $1^-$ state at excitation energy of 5.79 
MeV with a width of 28 eV. Reproducing the known value with this technique is an ideal 
opportunity to verify the method.

\item[Method]
The 1$^-$ state at 5.79 MeV is studied using the $\alpha$-transfer reaction 
$^{16}$O($^6$Li,$d$)$^{20}$Ne at sub-Coulomb energies.
 
\item[Results]
The partial $\alpha$ width for the $1^-$ state at excitation energy of 5.79 MeV is 
extracted and compared with the known value, allowing the accuracy of the method to be evaluated.

\item[Conclusions]
This study demonstrates that extracting the Asymptotic Normalization Coefficients using
sub-Coulomb $\alpha$-transfer reactions is a powerful tool that can be used to determine
the partial $\alpha$ width of near threshold states that may dominate  astrophysically 
important nuclear reaction rates.
\end{description}
\end{abstract}

\pacs{}

\maketitle

Nuclear reaction rates that involve $\alpha$-particles are often key nuclear physics inputs required for stellar models. The prime example is the $^{12}$C($\alpha$,$\gamma$) reaction, important in many astrophysical scenarios. Yet, direct measurements of the $\alpha$ induced reaction cross sections at energies that are relevant for stellar environments have not been possible. The product of the reaction cross section and the Maxwell-Boltzmann energy distribution for $\alpha$-particles in a stellar environment defines the energy range at which the specific reaction is most efficient. This energy range, known as the Gamow window, is typically far below the Coulomb barrier, where the Coulomb repulsion dominates and therefore the nuclear reaction cross section is very small and drops exponentially with energy. Since the cross section is often too small to be measured directly we are forced to rely on extrapolation of measurements done at higher energies down to the energies of interest. However, the reliability of 
these 
extrapolations is handicapped by the unknown nuclear structure of the systems involved. For example, direct measurements of the $^{12}$C($\alpha$,$\gamma$) reaction cross section have been performed only down to 900 keV in the center of mass frame (c.m.), while the Gamow window for the helium burning stage is around 300 keV.  The extrapolation is strongly affected by the sub-threshold states in $^{16}$O. Indirect methods can be used to constrain the properties of these resonances and therefore reduce the uncertainties related to low energy extrapolations. One of such methods is the $\alpha$-transfer reaction performed at sub-Coulomb energy, suggested by C. Brune, et al., \cite{Brun99}. By measuring the $\alpha$-transfer reaction cross section at energies low enough to be below the Coulomb barrier in both entrance and exit channels the dependence of the result on the optical model parameters is significantly reduced. Moreover, if the asymptotic normalization coefficients (ANCs) are extracted instead of the Spectroscopic Factors (SFs) then the 
dependence on the 
shapes of the $\alpha$-cluster form factors and the number of nodes of the cluster wave function is also eliminated. Therefore, this technique yields an almost model independent result, as long as the peripheral direct reaction mechanism dominates. Only three experiments that use this approach have been performed so far \cite{Brun99,John06,John09}. This is partially due to experimental difficulties in dealing with low recoil energies, but also due to more fundamental objections related to knowledge of the reaction mechanism. The main goal of this letter is to provide direct and unambiguous verification of a technique that has the potential to eliminate large uncertainties in the determination of astrophysically important reaction rates.

The key to proving this technique is the choice of a specific case that can serve as its verification. The nearly ideal opportunity to test the sub-Coulomb $\alpha$-transfer approach is provided by the 1$^-$ state at 5.79 MeV in $^{20}$Ne.
It is a purely $\alpha$-cluster state with partial $\alpha$ 
width close to the Single Particle (SP) limit.
This state is above the $^{20}$Ne $\alpha$-decay threshold by 1.06 MeV. Its natural width is known with good accuracy to be 28(3) eV \cite{Till98_20}. The natural width is also equal to the partial $\alpha$ width, since this state decays exclusively by $\alpha$ emission to the ground state of $^{16}$O (its partial $\gamma$-width is negligible). The ANC extracted from the $^{16}$O($^6$Li,$d$) reaction can be directly related to the partial $\alpha$ width \cite{Mukh99}, which can be compared to the directly measured natural width of the state, using the equation:
\begin{equation}\label{eq:PartialWidth}
    \Gamma_{\alpha} =P_l(kR)\frac{W^2_{-\eta,l+1/2}(2 k R)}{\mu R}(C^{^{16}\text{O}}_{\alpha^{12}\text{C}})^2, 
\end{equation}
where $P_l$ is the penetrability factor, $R$ is the channel radius, $k = \sqrt{2\mu \epsilon}$ is the wave number, with reduced mass $\mu$ and binding energy $\epsilon$, $W$ is the Whittaker function, $\eta$ is the Sommerfeld parameter and $(C^{^{16}\text{O}}_{\alpha^{12}\text{C}})^2$ 
is the ANC.

The experiment was performed at the John D. Fox superconducting linear accelerator facility at Florida State University.
It was crucial for this experiment to be performed at sub-Coulomb 
energies to avoid any dependence of the results on the entrance and exit channel optical 
potential parameters. 
Therefore, inverse kinematics was used to reach lower energies in the center of mass frame.
The $^{16}$O beam was produced by an FN Tandem Van de Graaff accelerator using a 
SNICS-II cesium-sputter ion source. 
The $^6$Li 
targets were prepared under vacuum and transported to the chamber in a
vacuum container in order to prevent oxidation. 
Several $^6$Li targets of thicknesses of about 50 $\mu$g/cm$^2$ were used. 
Since the $^6$Li targets have to remain under vacuum 
their thickness measurements have to be performed in-situ  
by using $^{16}$O+$^6$Li elastic scattering data.

The identification of the reaction products was performed using two $\Delta E$-$E$ telescopes designed specifically for the low energy $\alpha$-transfer reaction measurements. Each $\Delta E$-$E$ telescope is composed of four pin diode 2$\times$2 cm$^2$ silicon detectors and one position sensitive proportional counter in front of them. These components are contained in a box filled with P10 gas (10\% methane and 90\% Ar gas mixture). A Kapton foil of 7.5 $\mu$m thickness was used to separate the gas filled volume of the box from the vacuum of the scattering chamber. The scattering angle of the recoils is measured using the position of the hit
in the proportional counter.

\begin{figure}[t!]
\begin{center}\includegraphics[scale=0.3]{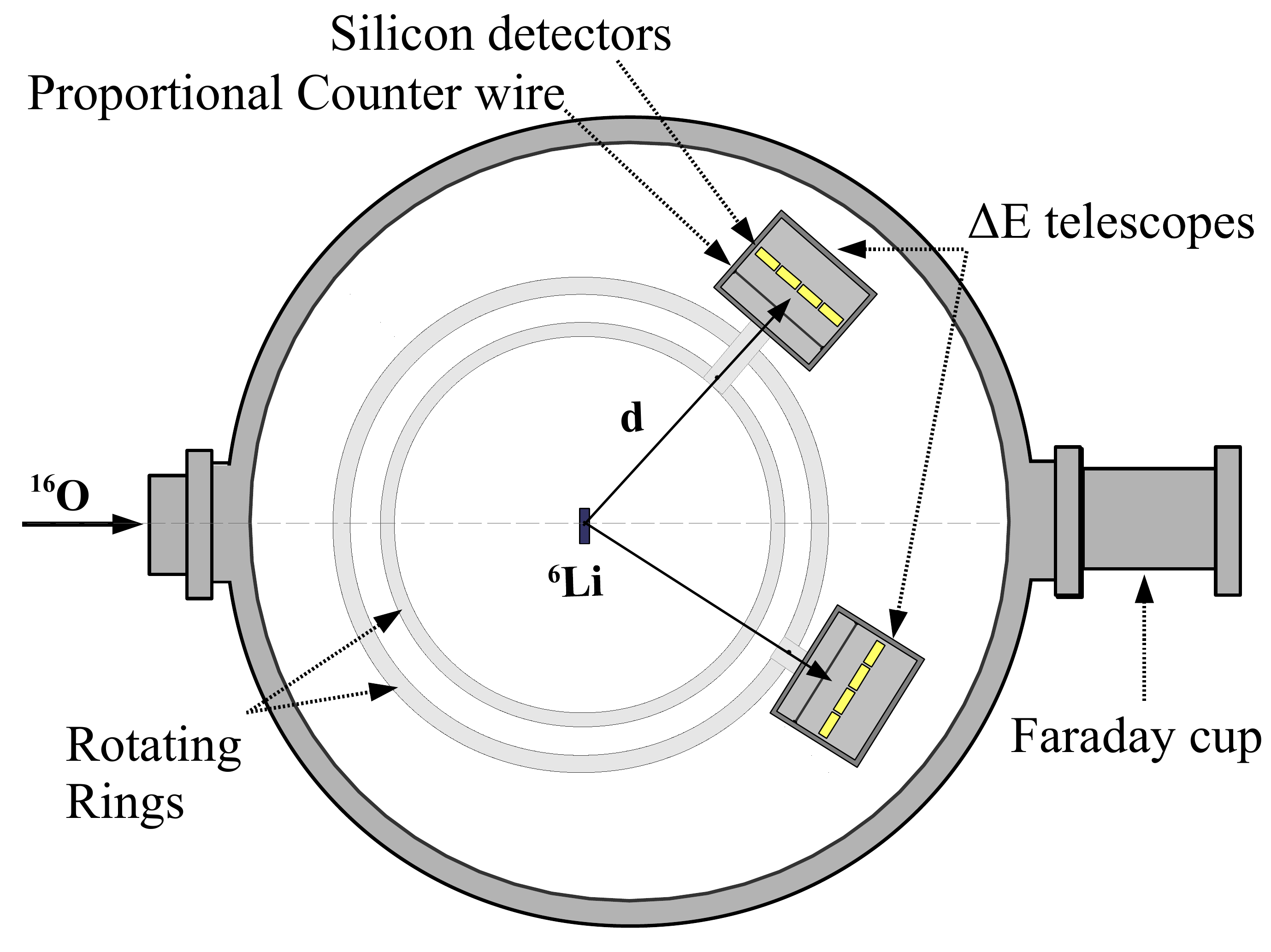}\end{center}
\caption{\label{fig:ANC_chamber} Top view of the experimental setup for ANC measurements.}
\end{figure}

The two $\Delta E$-$E$ telescopes were mounted on remotely controlled rotating rings and placed on both sides of the beam as shown in Fig. \ref{fig:ANC_chamber}. The pressure of the gas inside the detector box was optimized depending on the recoil to be measured. A pressure of 150 Torr was used for the measurements
of the deuterons and 50 Torr for the elastically backscattered $^6$Li.
The intensity of the incoming beam was measured using a Faraday cup placed at the end of the scattering chamber (Fig. \ref{fig:ANC_chamber}).

The absolute normalization of the cross section was determined from $^{16}$O+$^6$Li elastic scattering by measuring the backscattered $^6$Li ions. The elastic scattering cross section was calculated using the
code \textsc{fresco} (version FRES 2.9) \cite{thom88} with an optical potential 
obtained from \cite{Vine84}. 
For a beam energy below 13 MeV the scattering cross section is equal to Rutherford cross section at all but the most backward angles. But even at the scattering angle 180$^\circ$ in the center of mass, the cross section is grater than 70\% of Rutherford.

The elastic scattering data were measured between the production runs for each target and no statistically significant change in the normalization factor was observed, implying that the $^6$Li content of the targets was constant over time. However, it was observed 
that long exposure of the target to the low energy beam 
produced an energy shift of the $^6$Li peak to lower values as the run progressed. This was attributed to carbon buildup (from vacuum pumps and walls of the beam line) 
on the surface of the target making a slight change 
in the beam energy (due to energy loss in the carbon layer) and therefore making a shift in the $^6$Li peak. Normally this is not a problem 
because the $^6$Li content of the target does not change. However, since this experiment was performed at sub-Coulomb energy and the reaction cross section is very sensitive to the beam energy this beam energy loss must be determined.
To calculate the increment on the target thickness due to 
carbon buildup, elastic scattering data were taken when a target was used for the first time and every 2 hours of use after that. Any significant carbon buildup that increases the target thickness can be detected by an energy shift of the $^6$Li elastic peak after exposure.
This effect of beam energy reduction due to carbon buildup over time, while relatively small due to frequent target change (every 5-10 hours), was taken into account in the DWBA analysis by a corresponding reduction of the beam energy in the calculations. The typical beam energy 
in the middle of the target after the carbon build up
corrections are taken into account is 12.57 MeV.

\begin{figure}[t!]
\begin{center}\includegraphics[scale=0.44]{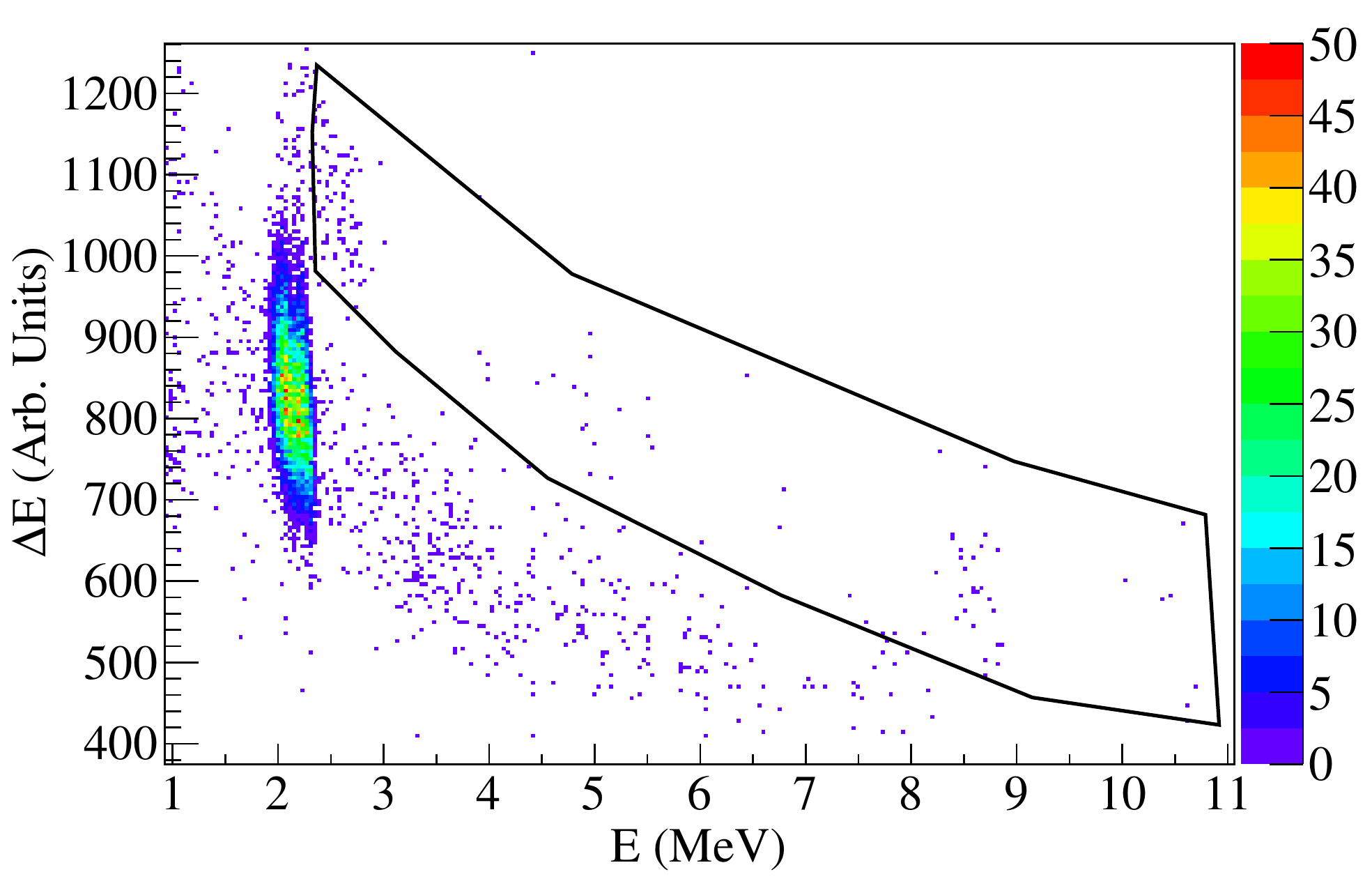}\end{center}
\caption{\label{fig:EdE} $\Delta E$ vs $E$ plot showing the deuterons cut for beam energy of 
12.57 MeV for the pin detector at 21$^\circ$ in the laboratory frame.}
\end{figure}

Deuterons were identified using a $\Delta E$ vs $E$ spectrum. Figure
\ref{fig:EdE} shows the $\Delta E$ vs $E$ spectrum for a pin detector 
at 21$^\circ$. A clear separation between the protons and deuterons is observed, except for a region at 2.1 MeV, where a strong proton background is observed. These protons are due to hydrogen content
in the target that produces elastically backscattered protons. This background restricts the $1^-$ state angular distribution to larger c.m. angles (five out of seven measured).

\begin{figure}[t!]
\begin{center}\includegraphics[scale=0.44]{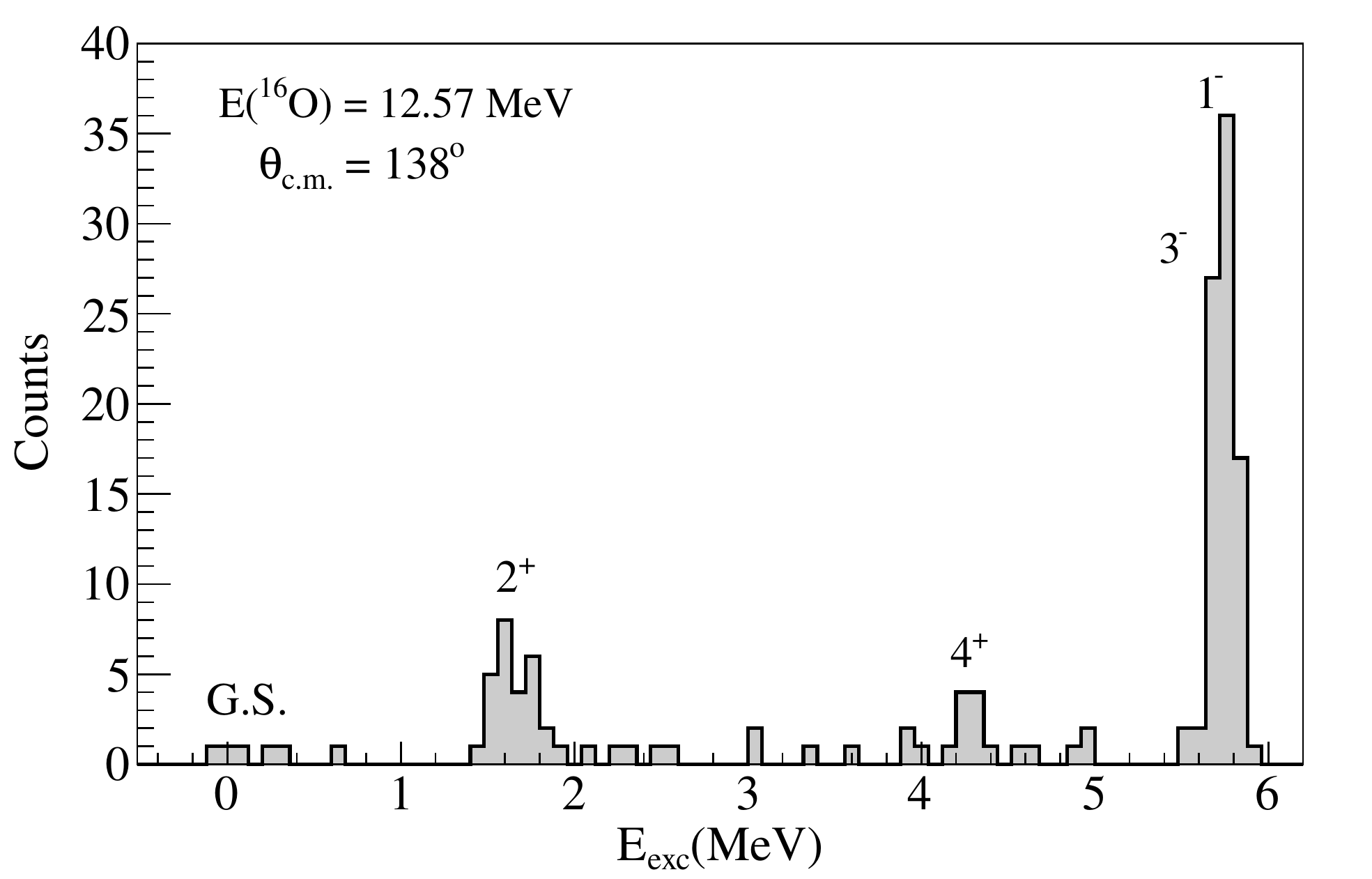}\end{center}
\caption{\label{fig:Exc_13} $^{20}$Ne excitation energy reconstructed from deuterons from the $^6$Li($^{16}$O,$d$) 
reaction at $\theta_{c.m.}$=138$^\circ$. The energy of $^{16}$O beam in the middle of the target is 12.57 MeV.} 
\end{figure}

\begin{figure}[ht]
\begin{center}\includegraphics[scale=0.44]{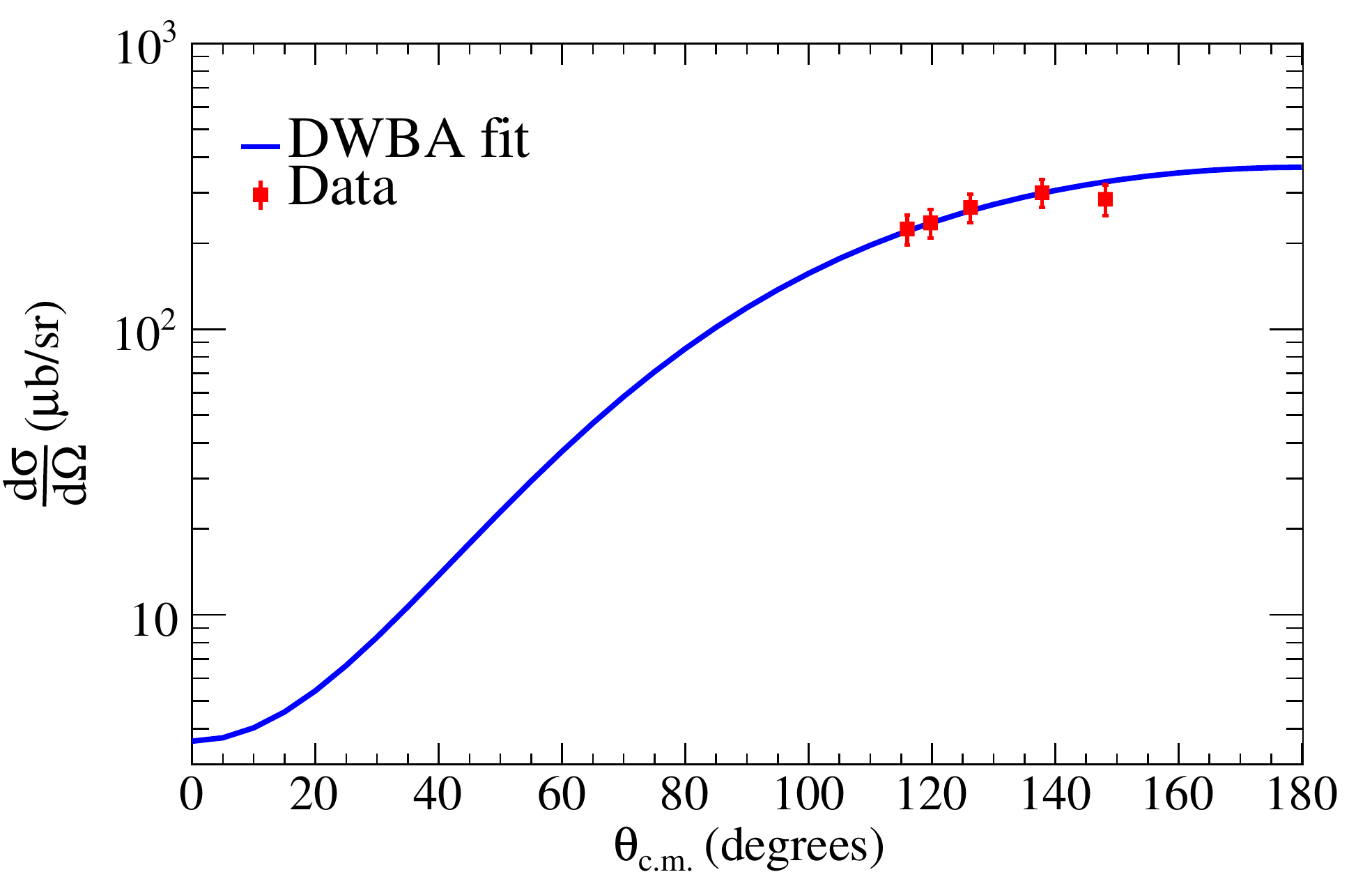}\end{center}
\caption{\label{fig:Cross_s_13} Angular distribution of the $1^-$(5.79 MeV) state in 
$^{20}$Ne and DWBA fit for E($^{16}$O)=12.57 MeV.}
\end{figure}

The $^{20}$Ne excitation energy reconstructed from deuterons measured at $\theta_{c.m.}=138^{\circ}$ is shown in Fig. \ref{fig:Exc_13}. The $x$-axis corresponds to the excitation energy in $^{20}$Ne. All low lying states in $^{20}$Ne are clearly observed, except for the unnatural parity 2$^-$ state at 4.97 MeV that cannot be populated in direct, single-step $\alpha$ transfer. The measurements are essentially background free at this energy. The 3$^-$ state at 5.62 MeV cannot be resolved from the 1$^-$ at 5.79 MeV, but the cross section to populate this state is very small (see below) and we attribute all counts observed around 5.8 MeV to the 1$^-$ state.    
The angular 
distribution for the 1$^-$ state is shown in Fig. \ref{fig:Cross_s_13}.

The theoretical analysis of the cross section is done using the finite range
DWBA approach via the computer code \textsc{fresco}. The calculations were performed using a 
finite range transfer including a full complex remnant term. 
The potentials used in the DWBA calculations are 
given in Table \ref{tab:20NePot}. 
\begin{table*}[ht]
\centering
\caption{\label{tab:20NePot}Parameters of the potentials used in DWBA calculations for $^{20}$Ne. For 
$d$+$\alpha$ and $\alpha$+$^{16}$O, $V_0$ was fitted to reproduce the binding energies
of $^6$Li and $^{20}$Ne, respectively. All the radii $r_x$ are given such that
$R_x=r_x(A_p^{1/3}+A_t^{1/3})$.}
\tabcolsep=0.18cm
\begin{tabular}{ccccccccccccc}
\hline
\hline
 Channel & $V_0$ & $r_v$ & $a_v$ & $W$  & $W_s$ &$r_w$ & $a_w$ & $r_c$ &$V_{so}$& $a_{so}$& 
 $r_{so}$ & ref. \tabularnewline
    & (MeV) & (fm)  &  (fm) & (MeV) & (MeV) & (fm) & (fm) & (fm)  & (MeV) & (fm) & (fm) & \tabularnewline  	
\hline
 $^6$Li+$^{16}$O& 159 &	0.71 & 0.83 & 4.26	& - & 1.40 & 0.81 & 1.25 & - &- &- & \cite{Vine84}\tabularnewline
 $d$+$^{20}$Ne    & 105 &	0.70 & 0.86 & - & 24  & 0.97 & 0.65 & 1.25 & 6 & 0.86 & 0.70& \cite{John06} \tabularnewline
 $d$+$^{16}$O	& 79.5 & 0.83 & 0.8 & 10 & - & 0.83 & 0.8 & 1.25 & 6 & 0.8 & 0.83&\cite{John06} \tabularnewline
 $d$+$\alpha$	& -   & 0.70 & 0.65 &   &  &  &  &  &  &  & & \cite{Kubo72}\tabularnewline
 $\alpha$+$^{16}$O	& - & 0.77 & 0.8 &  &  &  &  &  &  &  & & \tabularnewline
\hline
 \hline
\end{tabular} 
\end{table*}
The radius is defined as $R_x=r_x(A_t^{1/3}+A_p^{1/3})$.
The parameters for the  $^6$Li+$^{16}$O optical potential are based on 
\cite{Vine84} where $^6$Li+$^{12}$C elastic scattering was studied in the
energy range from 4.5 to 50.6 MeV. The 
$d$+$^{20}$Ne and $d$+$^{16}$O optical potential parameters are the
same as those used in \cite{John06}. For the $^6$Li formfactor, an 
$\alpha$+$d$ configuration was assumed to have $R_v=1.9$ fm and
$a=0.65$ fm. These parameters were obtained from \cite{Kubo72}.
$V_0$ was fitted to reproduce the binding energy of $^6$Li. The final
results are almost independent of the choice of potential parameters for
this sub-Coulomb $\alpha$-transfer reaction (see discussion below).

\begin{figure}[ht]
\begin{center}\includegraphics[scale=0.45]{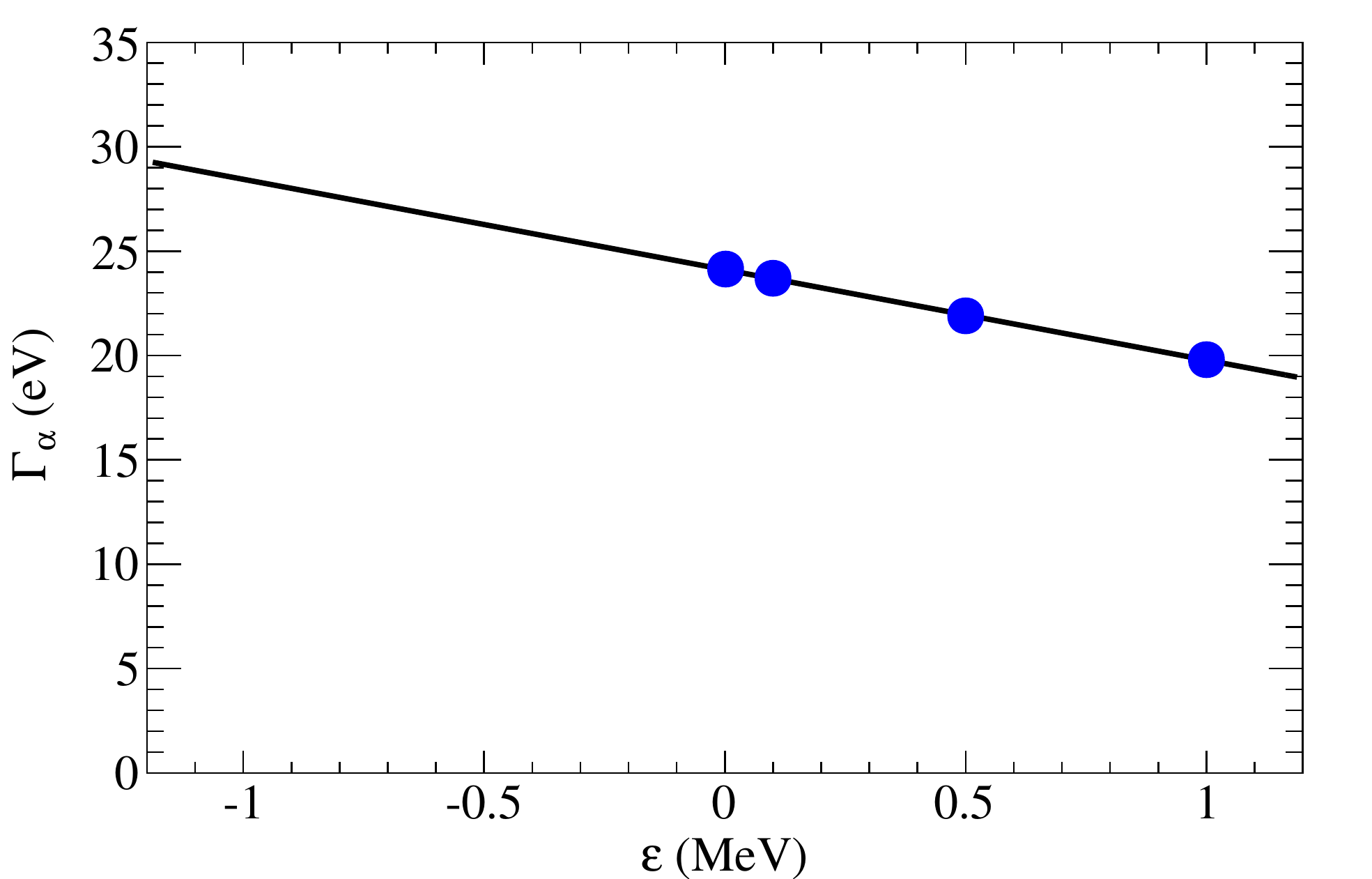}\end{center}
\caption{\label{fig:ANCvsBE} Partial $\alpha$ width as a function
of binding energy for the 1$^-$ (5.79 MeV) state in $^{20}$Ne.}
\end{figure}

The existing DWBA codes are designed for calculating transfer cross
section into the bound states and since the 1$^-$ at 5.79 MeV is an
unbound state an artificial binding energy was used in the calculations.
The fit shown in Fig. \ref{fig:Cross_s_13} is obtained using a binding
energy of 0.1 MeV. The value of the ANC and partial $\alpha$ width
calculated from it, using Eq. \ref{eq:PartialWidth}, depend on the choice of
binding energy so that the partial $\alpha$ width for different 
binding energies was calculated and a nearly linear dependence on
the binding energy was found as shown in Fig. \ref{fig:ANCvsBE}. Linear
extrapolation allows the partial $\alpha$ width for the correct 
binding energy of -1.06 MeV for this unbound state to be determined.  
The Whittaker function and penetrability factor are calculated using a
channel radius of $R=5.1$ fm (dependence of the final result on this
parameter is discussed below). The final result obtained for the partial 
$\alpha$ width for the unbound $1^-$ state at excitation 
energy of 5.79 MeV in $^{20}$Ne is   
$\Gamma_{\alpha}=29(6)\hspace{0.2cm}\text{eV}$. 
This result is in excellent agreement 
with the known value of 28(3) eV \cite{Till98_20}. The validity and accuracy of
the ANC method is thus verified. Evaluation of a possible contribution from the compound
nucleus reaction mechanism (which can potentially limit the applicability of the
method) and a contribution to the $1^-$ state yield from the unresolved 3$^-$ state at 5.62 MeV are discussed in the next two paragraphs.  

The Compound Nucleus (CN) contribution was estimated using the computer
code EMPIRE (version EMPIRE-3.2) \cite{Herm07}. The calculated total cross section for the population of
the 1$^-$ state at 5.79 MeV is 6.5 $\mu$b. Assuming a uniform angular
distribution yields a corresponding differential cross section of  0.5 $\mu$b/sr which is to be compared
to the experimental 200 - 300 $\mu$b/sr cross section. The total CN cross
section for the unresolved 3$^-$ state at 5.62 MeV is 14 $\mu$b, which then
corresponds to $\sim$1 $\mu$b/sr. The EMPIRE calculations can be verified using the 2$^-$ state yield at 4.97 MeV since this unnatural parity state  can only be populated by the CN or multistep mechanisms. To obtain an upper limit of the CN cross section it was assumed that all the counts (3) seen in Fig. \ref{fig:Exc_13} at 4.97 MeV come from the CN mechanism. The experimental cross section for the 2$^-$ state (although we can only attribute a few counts to this state) is $\sim$15 $\mu$b/sr. The EMPIRE calculations predict 128 $\mu$b total cross section, which corresponds to $\sim$10 $\mu$b/sr, a value consistent with experiment. Therefore, we conclude that the CN mechanism cannot contribute more than 1\% to the observed cross section. However, the 1$^-$ at 5.79 MeV is a highly clustered state with partial $\alpha$ width close to the single particle limit whereas for states that have $\alpha$-cluster strength at the level of few percent the CN mechanism may become an important limiting factor of the method. 

The contribution of the unresolved 3$^-$ state at 5.62 MeV was evaluated using the relative $\alpha$ spectroscopic factor (normalized to unity for the g.s.) measured in \cite{Anan79}, where the $\alpha$-transfer reaction $^{16}$O($^6$Li,$d$) at bombarding energies of 20, 32 and 38 MeV was studied. 
The relative $\alpha$ strength obtained in \cite{Anan79} for the 1$^-$(5.79 MeV) and 3$^-$(5.62 MeV) states was 0.51 and 0.06, respectively. 
We calculated the cross section for population of the 1$^-$ and the 3$^-$ with unity $\alpha$-SF for both using FRESCO code and the potentials given in Table \ref{tab:20NePot}. Then we scaled the
3$^-$ cross section by a factor of 0.06/0.51=0.12. The resulting ratio between the cross section for the population of the 1$^-$ and 3$^-$ states is 0.03. Therefore, the 3$^-$ state contributes 3\% to the cross section. Subtracting this contribution from the experimental cross section would make the partial $\alpha$ width for the 1$^-$ state equal to 28(6) eV.

To determine the precision of the extracted partial $\alpha$ width several factors are taken into account. The statistical uncertainty related to the number of events in the measurement is 12\%. The normalization uncertainty is calculated by using slightly different energies as well as measuring the target thicknesses with two different beams. 
For some of the targets the thickness was also measured using an $^{16}$O beam at 10 MeV and a $^{12}$C beam at 9 MeV to study the dependence on the energy and the beam used. Assuming different interaction places in the target (instead of in the middle of the target) gives small variation in the beam energy at the moment of interaction. The calculated normalization uncertainly is 10\%.
For the DWBA analysis it was found that using different parameters for the potentials produces variations of less than 10\%. In fact calculations with no optical potentials (only Coulomb) gives a difference of about 13\%. 
The number of nodes used in this calculation for the partial width is four. 
Using one less and one extra number of nodes gives 8\% variation in the result. Variations of the partial $\alpha$ width associated with different values of channel radius (varied from 4.7 to 5.5 fm) is less than 9\%. The combined total uncertainty for the partial width of the 1$^-$ state at 5.79 MeV is determined to be 22\%.

In summary, we have verified that an $\alpha$-transfer reaction  performed at sub-Coulomb energies can produce an accurate and model independent determination of the asymptotic normalization coefficients (ANCs) 
of the near-threshold resonances and sub-threshold states and then these ANCs can be used to constrain key astrophysical reaction rates.
The precision that can be achieved in these experiments is limited by the influence of the optical model potentials, which can be mitigated by reducing the beam energy and going deeper below the Coulomb barrier. The accuracy is limited by the contribution of the compound nucleus mechanism to the reaction cross section and is the irreducible
limitation. 
However, it is expected to be small in most realistic situations and was shown to contribute less 
than 1\% to the ANC of the 1$^-$ state in $^{20}$Ne measured in this work. The results presented here
validate the sub-Coulomb $\alpha$-transfer method which can be used to constrain the contribution 
of the near-threshold resonances and sub-threshold states to the $\alpha$ induced reaction rates. 
The important point is that the method is not only applicable for the experiments with stable beams, 
but also can be used with good quality (reaccelerated) low energy rare isotope beams.

\begin{acknowledgments}
The authors would like to acknowledge the financial support provided
by the National Science Foundation under grant No. PHY-456463,
the U.S. Department of Energy under contract
No. DE-FG03-93ER40773 and the Welch Foundation (Grant No.: A-1853).
\end{acknowledgments}

\bibliography{myrefs}

\end{document}